\documentstyle[12pt]{article}

\def\sg#1{\hbox{\sf#1}}
\newcommand{\be}{\begin{eqnarray}}
\newcommand{\ee}{\end{eqnarray}}
\newcommand{\bi}{\bibitem}
\newcommand{\la}{\label}

\newcommand{\benl}{\begin{eqnarray*}}
\newcommand{\eenl}{\end{eqnarray*}}
\font\twelve=cmbx10 at 15pt
\font\ten=cmbx10 at 12pt

\renewcommand{\thefootnote}{\fnsymbol{footnote}}
\begin{document}

\begin{titlepage}

\begin{center}

{\ten Centre de Physique Th\'eorique\footnote{Unit\'e Propre de
Recherche 7061} - CNRS - Luminy, Case 907}

{\ten F-13288 Marseille Cedex 9 - France }

\vspace{1 cm}

{\twelve ON LONG-RANGE ORDER IN LOW-DIMENSIONAL LATTICE-GAS
MODELS OF NEMATIC LIQUID
CRYSTALS}

\vspace{0.3 cm}
\setcounter{footnote}{0}
\renewcommand{\thefootnote}{\arabic{footnote}}

{\bf N. ANGELESCU\footnote{Department of Theoretical Physics,
Institute for Atomic Physics,
PO Box MG-6,
Bucharest (ROMANIA)}, S. ROMANO\footnote{Dipartimento di Fisica
``A. Volta'', Universit\`a,
via A. Bassi 6,
I-27100 Pavia (ITALY)} and
V.A. ZAGREBNOV\footnote{and Universit\'e d'Aix-Marseille II}}

\vspace{1.5 cm}
{\bf Abstract}

\end{center}

The problem of the orientational ordering transition for
lattice-gas models
of liquid crystals is discussed in the low-dimensional case
$d=1,2$.
For isotropic short-range interactions, orientational long-range
order at finite temperature
is excluded for any packing of molecules on the lattice $Z^d$; on
the other hand, for reflection-positive long-range isotropic
interactions, we prove existence of an orientational ordering
transition for high packing ($\mu > \mu_0$) and low temperatures
($\beta > \beta_c(\mu)$).

\vspace{2 cm}

\noindent PACS numbers 05.50, 61.30, 64.70

\bigskip

\noindent October 1994

\noindent CPT-94/P.3087

\bigskip

\noindent anonymous ftp or gopher: cpt.univ-mrs.fr

\end{titlepage}

\parindent 0cm
\section{~~~}
The problem of long-range orientational order in models mimicking
liquid crystalline (usually nematic) behaviour has been discussed
at the rigorous level and in the framework
of different interaction models, see, e.g, Ref.
\cite{r01,r02,r03,r04}.

In the following, we shall be considering cylindrically symmetric
molecules, whose centres of mass belong to a $d-$dimensional space
($x \in Z^d$, {\it lattice models}, or $x \in R^d$, {\it continuum
models}), and whose orientations are defined by $m-$component unit
vectors $u \in S^{(m-1)} \subset R^m,~m \ge 2$.

For low-dimensional {\it lattice models } (i.e. dimension $d=1,2$)
with isotropic short-range interactions, the Mermin-Wagner theorem
entails
absence of an
orientational ordering transition taking place at finite
temperature;
a similar result was proven by Romerio for {\it continuum} fluids
\cite{r01}.
On the other hand, a two-dimensional {\it lattice} model with
anisotropic interactions restricted to nearest neighbours can
produce an ordering transition (also of the nematic type), see,
e.g. Ref. \cite{r10}, Ref. \cite{rgeor}, Ch. 9, Example 9.22(2).

The proof of the
existence of orientational order
for {\it continuum} liquids
is a rather complicated problem.
In this connection we would like to mention Ref. \cite{r05}, where
the existence of a ferromagnetic phase transition is proven for a
{\it continuum} fluid
($d \ge 2$) of classical particles
carrying {\it Ising}-like spins
and having suitable magnetic and non-magnetic interactions.
The proof is based on the FKG- and GHS-inequalities, which make it
possible
to find a lower bound for the magnetization of the fluid in terms
of that of a suitable spin system on a lattice; no such
inequalities are known for $m \ge 3$. 

The aim of the present note is to
prove
the existence of orientational ordering (at finite temperature) in
low-dimensional {\it lattice-gas} models ($d=1,2$) \cite{r04}
with isotropic long-range interactions
of nematic symmetry, which
therefore do not satisfy the hypotheses
of the Romerio theorem \cite{r01}.

\parindent0cm
\section{~~~}
In order to make
contact with (nematic) liquid crystals, and following the standard
phenomenology \cite{r06},
we consider centrosymmetric molecules (symmetry $D_{\infty h}$),
and define the matrix
\be
Q^{\alpha \beta}=
\bigl( u^{\alpha} u^{\beta} - \frac{1}{m} \delta_{\alpha \beta}
\bigr),~ \alpha,\beta=1,2,\ldots m;
\label{e01}
\ee
then translationally invariant
interactions between molecules can be expressed as \be
V_{xy}=-a(|x-y|)\,Tr(Q_x \cdot Q_y)~+~g(|x-y|),~x,y \in R^d.
\label{e02}
\ee
The Hamiltonian of the lattice-gas caricature of the nematic liquid
crystal has the form \cite{r04}:
\be
H_{\Lambda}(\tilde{n},\tilde{Q})=-\frac{1}{2} \sum_{(x,y)} n_x n_y
J_{xy} Tr(Q_x \cdot Q_y) + \frac{1}{2} \sum_{(x,y)} n_x n_y I_{xy}
- \mu \sum_{x\in \Lambda}n_x,
\label{e03}
\ee
where $(x,y) \equiv \{ x,y \in \Lambda : x \ne y \}$. In this
context we can assume $a(0)=g(0)=0$.

So, molecules live on the sites of the cubic sublattice $\Lambda
\subset Z^d$ with periodic boundary conditions, i.e. \be
J_{xy}=\sum_{\{ z \in Z^d:z=y(\bmod \Lambda) \} } a(|x-z|),~
I_{xy}=\sum_{\{ z \in Z^d:z=y(\bmod \Lambda) \} } g(|x-z|).
\nonumber
\ee
A configuration is specified by a set of occupation numbers $\{
n_x=0,1;~x \in Z^d\}$ and, for all $\{ x \in Z^d:n_x=1 \}$, by the
configuration $Q_x$ of the molecule at $x \in Z^d$. The
corresponding finite-volume Gibbs state is defined by
$$
\langle f \rangle_{\Lambda}(\beta,\mu)=
\bigl[\Xi_{\Lambda}(\beta,\mu)\bigr]^{-1}$$
\be \sum_{\{n_x=0,1;~x \in
\Lambda \}}
\int \prod_{x\in \Lambda} (d \nu(Q_x))^{n_x} exp[-\beta
H_{\Lambda}(\tilde{n},\tilde{Q})] f(\tilde{n},\tilde{Q}),
\label{e04}
\ee
where $\Xi_{\Lambda}$ is the partition function, $d \nu$ is the
$O(m)-$invariant probability measure induced by the Haar measure on
the unit sphere in $R^m$, see Eq.\ (\ref{e01}), and $\tilde{n}
\equiv \{ n_x;~x \in \Lambda \}$, $\tilde{Q} \equiv \{ Q_x;~x \in
\Lambda\}$.

The chemical potential $\mu$ governs the concentration (mean
density) $\rho(\beta,\mu)=\langle n_x \rangle _{\Lambda}$, of
molecules on the lattice at the temperature $\beta ^{-1}$. The
interaction term $J_{x y} Tr (Q_x \cdot Q_y)$ involves both
positional and orientational degrees of freedom, and possibly
produces orientational order, whereas
$I_{xy}$ mimics a direct interaction between them (positional
order).

Let $a(|x-y|)$ denote a short-range interaction ($SR$,
i.e. finite-ranged or decreasing at least exponentially), or a
long-range one
with asymptotic behaviour
$a(|x-y|) \sim |x-y|^{-(d+\sigma)}$, for $|x-y| \rightarrow
\infty$, $\sigma > 0$. 
Then, in the limit $q \rightarrow 0$,
the lattice Fourier transform
of $J_{x y}$
is given by
\be
\hat{J}(0)-\hat{J}(q) \simeq
\left\{ \begin{array}{ll}
c |q| ^{\sigma},&0 < \sigma < 2
\\
c |q|^2,&\sigma \ge 2,~or~SR
\end{array}
\right.
\label{e05}
\ee
where

$q \in \Lambda^*\equiv \bigl\{ q^{\alpha}=\frac{2
\pi}{|\Lambda|^{1/d}} m_{\alpha},~
m_{\alpha}=0,\pm 1,\ldots,\pm \bigl( \frac{|\Lambda|^{1/d}}{2}-1
\bigr), +\frac{|\Lambda|^{1/d}}{2},~\alpha=1,2,\ldots d \bigr\}$.

Then,
since this interaction is $O(m)-$invariant, we have 
the following Proposition \`{a} la Mermin-Wagner, due to
\cite{rlast}.

\underline{Proposition 2.1}

Let $d=1,2$, $\sigma \ge 2$ (or let $a(|x-y|)$ define a $SR$
interaction), and let $g(|x-y|)$
be such that
\benl
\sum_{y \in Z^d} |g(|x-y|)|\,|x-y|^2 < +\infty. \eenl

Then there is no orientational order at finite temperature, i.e.
\be
P(\beta,\mu) \equiv \lim_{|x-y| \rightarrow \infty} \lim_{\Lambda
\uparrow Z^d}
\langle Tr(D_x \cdot D_y) \rangle_{\Lambda}= 0, \label{e06}
\ee
Here $D_x \equiv n_x \cdot Q_x$.

\underline{Proof}

For any fixed configuration $\tilde{n}$ the limiting Gibbs state
$\langle - \rangle(\beta,\tilde{n})$ is $O(m)-$invariant. In this
case, the proof developed in \cite{rlast} carries through verbatim
for the $O(m)-$invariant interaction
defined by (\ref{e02}) and corresponding to $U_{xy}(Q_x,Q_y)$ in
Pfister's notation \cite{rlast}.
Since the conditions both on $I_{xy}$ and $J_{xy}$ guarantee for
the Hamiltonian (\ref{e03}) the equivalence of ensembles (see Ref.
\cite{rmast,rnast}), one gets that the grand-canonical Gibbs state
\be
\langle f \rangle (\beta,\mu)=
\int d K(\mu,\tilde{n})
\langle f \rangle(\beta,\tilde{n})
\la{el01}
\ee
is also $O(m)-$invariant. Here $d K(\mu,\tilde{n})$ is the
Kac-transformation kernel \cite{rmast,rnast}. For any pure state
$\langle - \rangle(\beta,\mu)$, we get
\be
\lim_{|x-y| \rightarrow \infty} \langle Tr(D_x \cdot D_y) \rangle =
\langle TrD_x \rangle^2 = 0,
\la{el02}
\ee
where the last equality follows from the $O(m)-$invariance of the
state (\ref{el01}).~~~~$\Box$

The presence of ``holes'' ($n_x=0$) evidently disfavours the
orientational order on $\lim_{\Lambda \uparrow Z^d} M(\tilde{n})$.
It is natural to guess that the order parameter $P(\beta,
\tilde{n})$, and hence $P(\beta,\mu)$ are bounded by the order
parameter of the lattice without ``defects'', i.e. $\{ n_x=1;~x \in
Z^d\}$, which corresponds to $\mu \rightarrow +\infty$.
One can easily check this for particular cases of the lattice-gas
Ising system (where $Tr(Q_x \cdot Q_y)$ is substituted by $\tau_x
\cdot \tau_y$, $\tau_{x,y}=\pm 1$), or for the lattice-gas of plane
rotators, due to the GHS inequalities. Hence, for these systems,
the result (\ref{e06}) is a simple consequence of
``pure system domination'': $P(\beta, \mu) \le P(\beta,
\mu=+\infty)$, and of the Mermin-Wagner theorem for the regular
lattice $Z^d$, $d=1,2$.

This result can be futher strengthened:
for $d=1$ and $1 \le \sigma \le 2$,
the absence of orientational order
has also been proven for the lattice model(s) [Ref. \cite{rgeor},
Ch. 9, Comment 9.34; Theorem 14' in Ref. \cite{r07}; Refs.
\cite{radd2,radd3}].
Notice that Romerio's result \cite{r01} holds for $\sigma > 2$ but
for a {\it continuum} model.

\parindent 0cm
\section{~~~}
In contrast to the ``no-go'' Proposition 2.1, the proof of
existence of orientational order at finite temperature ($P(\beta,
\mu) > 0$) is a more delicate task.
We are able to do this for long-range interactions which are
reflection-positive with respect to reflections in planes without
sites \cite{r07,r08}.

{}From now on,
let $g(|x-y|)=0$ and
let $a(|x-y|)$ correspond to the long-range interaction \be
a(|x-y|)=b |x-y|^{-(d+\sigma)},~b > 0
\label{e07}
\ee
with $0 < \sigma < d$.
First we
note that,
for (\ref{e07}) the asymptotic form
of the Fourier transform (\ref{e05}) excludes a 
a ``no-go'' theorem \`{a} la
Mermin-Wagner.

\underline{Proposition 3.1}

For $d=1,2$ there exists a $\mu_0$
and, for every $\mu > \mu_0$ there exists a $\beta_c(\mu)$ such
that
$P(\beta,\mu) > 0$ for $\beta > \beta_c(\mu)$. This means that
every limiting Gibbs state $\langle - \rangle= \lim_{\Lambda
\uparrow Z^d} \langle - \rangle _{\Lambda}$
has long-range orientational order.

\underline{Proof}

The Proof is an adaptation
to our case (\ref{e01})
of the line of reasoning developed in
\cite{r04} for the case of the general matrix order parameter. By
(\ref{e01}) one has $Tr Q_x=0$, i.e. $Tr \langle D_x
\rangle_{\Lambda}= \langle Tr D_x \rangle_{\Lambda}=0$. Then , by
the $O(m)-$invariance of the state $\langle - \rangle_{\Lambda}$,
one can deduce that $\langle D_x \rangle_{\Lambda}=0$. Hence,
$P(\beta, \mu) > Tr(\langle D_x \rangle^2)= 0$ will really mean
existence of long-range orientational order.

Using the translational invariance of $\langle -
\rangle_{\Lambda}$, we get
\be
c_{\Lambda}=|\Lambda|^{-2}\sum_{(x,y)}
\langle Tr(D_x D_y)\rangle_{\Lambda}=
Tr\langle D_x^2\rangle_{\Lambda}-|\Lambda|^{-1} \sum_{p \in
\Lambda^* \setminus \{ 0 \} } Tr\langle \tilde{D}_p \cdot
\tilde{D}_{-p} \rangle_{\Lambda}, \label{e08}
\ee
where

$\tilde{D}_p= |\Lambda|^{-1/2} \sum_{x \in \Lambda}\exp(-i p x)
D_x$.

Thus, in order to prove that $P(\beta,\mu) > 0$, it suffices to
verify that
\be
c(\beta,\mu) \equiv \lim_{\Lambda \uparrow Z^d} \inf c_{\Lambda} >
0. \label{e09}
\ee
The upper bound on $Tr\langle \tilde{D}_p
\tilde{D}_{-p}\rangle_{\Lambda}$ for $p \in \Lambda^* \setminus \{0
\}$
results from the Infrared Bound \cite{r07,r08} \be
\langle Tr (\tilde{D}_p \tilde{D}_{-p})\rangle_{\Lambda} \le
\frac{const}{\beta [\hat{J}(0)-\hat{J}(p)]} \sim
|p|^{-\sigma},~for~ p \rightarrow 0 \label{e10}
\ee
which holds true for the reflection-positive interaction
(\ref{e07}), due to the chessboard estimate proving the gaussian
domination; see Refs. \cite{r10,r09} for a review. Therefore, the
sum over $\Lambda^* \setminus \{ 0 \}$ divided by $|\Lambda|$ (see
(\ref{e08})) can be estimated in the limit $\Lambda \uparrow Z^d$
{}from {\it above} by the integral \be
I_{d,\sigma}(\beta)= \frac{const}{\beta} \int_{[-\pi,+\pi]^d} d^d p
[\hat{a}(0)-\hat{a}(p) ]^{-1} < \infty,~0< \sigma <d.
\label{e11}
\ee
In order to get the lower bound of the first term of (\ref{e08}),
$Tr\langle D^2_x \rangle_{\Lambda}=\langle n^2_x \rangle_{\Lambda}
Tr(Q^2_x)=\frac{m-1}{m} \langle n_x \rangle_{\Lambda}$, we again
use the chessboard estimate \cite{r07,r08}: \be
\langle 1-n_x \rangle_{\Lambda} \le \big[ \langle \prod_{y \in
\Lambda} (1-n_y)\rangle_{\Lambda} \bigr]^{1/|\Lambda}.
\label{e12}
\ee
Then one gets
\be
\langle n_x \rangle_{\Lambda}
\ge 1 - \bigl[ \langle \prod_{y \in \Lambda} (1-n_y)
\rangle_{\Lambda} \bigr]^{1/|\Lambda|}. \label{e13}
\ee
According to (\ref{e04}) the last term in (\ref{e13}) is equal to
$[Z_{\Lambda}(\beta,\mu)]^{-1/|\Lambda|}$. To get a lower bound on
the partition function $Z_{\Lambda}(\beta,\mu)$, we choose a
$\overline{D} \in supp~\nu$ and a neighbourhood ${\cal N}_\epsilon$
of $\overline{D}$ such that
\be
Tr(D_1 \cdot D_2) > (1-\epsilon) Tr (\overline{D}^2) > 0,
\label{e14}
\ee
for some $\epsilon > 0$ and for every $D_1,D_2 \in {\cal
N}_{\epsilon}$. Then for configurations

$\Delta^{(\epsilon)} \equiv \bigl\{ D_x: D_x \in {\cal N}_{\epsilon},x
\in \Lambda\bigr\},$

we get by (\ref{e14})
that
$\{ n_x=1;~x \in \Lambda\}$ and
\be
-H_{\Lambda}(n,Q) \ge (1-\epsilon) Tr \overline{Q}^2 \sum_{(x,y) }
J_{xy} + \mu |\Lambda|=
[(1-\epsilon) Tr \overline{Q}^2 \hat{a}(0)+ \mu] \cdot |\Lambda|
\label{e15}
\ee
Therefore, upon restricting the integration in the partition
function to configurations $\Delta^{(\epsilon)}$, we obtain \be
[Z_{\Lambda}(\beta,\mu)]^{-1/|\Lambda|} \le \bigl[\nu({\cal
N}_{\epsilon})\bigr]^{-1}
\exp\bigl\{ -\beta \bigl[ (1-\epsilon)
Tr\overline{Q}^2 \hat{a}(0)+ \mu \bigr] \bigr\}, \label{e16}
\ee
and, as a consequence of (\ref{e13}), one gets \be
Tr\langle D^2_x \rangle_{\Lambda} \ge
\frac{2}{3} \bigl[ 1-\frac{1}{\nu({\cal N}_{\epsilon})} \exp \{
-\beta [ (1-\epsilon) Tr \overline{Q}^2 \hat{a}(0) + \mu ] \}
\bigr]
\equiv L(\beta,\mu)
\label{e17}
\ee
Combining (\ref{e11}) with (\ref{e17}), we obtain the following
estimate for (\ref{e09}):
\be
c(\beta,\mu) \ge L(\beta,\mu) - I_{d,\sigma}(\beta). \label{e18}
\ee
Now one immediately sees that for

$\mu > \mu_0 \equiv -(1-\epsilon) Tr( \overline{Q}^2) \hat{a}(0)$,

there exists a $\beta_c(\mu)$ such that

$L(\beta_c(\mu),\mu)=I_{d,\sigma}(\beta)$,

where
$0 < \sigma < 1,~d=1$ and $0< \sigma < 2,~d=2$. Hence, for $\beta
\\beta_c(\mu > \mu_0)$ we get $c(\beta,\mu) > 0$.
According to Eqs.\ (\ref{e08}), (\ref{e09}), this means that
$P(\beta,\mu)>0$ in the named $(\beta,\mu)$ domain, i.e. in any
pure limiting Gibbs state $\langle D_x \rangle= \langle Q_x \rangle
\neq 0$: thus the $O(m)$ symmetry is broken, which means long-range
orientational order.~~~~$\Box$

Notice that the sign of $a(|x-y|)$ plays an important role in the
existence theorem, but not in its absence counterpart.

Moreover,
the following corollaries can be obtained, by the same line of
reasoning as in the preceding propositions.

\underline{Corollary 1.}

Consider the ferromagnetic (lattice gas) counterpart of the present
model,
i.e. whose orientation-dependent two-body interaction reads
\be
V_{xy}=-a(|x-y|)\,(u_x \cdot u_y).
\label{elast01}
\ee
Then, under the same hypotheses on $a(|x-y|)$ and $g(|x-y|$, one
can prove absence or existence of an ordering transition,
respectively. 

\underline{Corollary 2.}

Let $m=3$, and let the
orientation-dependent two-body interaction have the general form
\be
V_{xy}=- a(|x-y|) P_L(u_x \cdot u_y),
\la{elast02}
\ee
for an arbitrary positive integer $L$; owing to the addition
theorem for spherical harmonics, the Legendre polynomial can be
written as \be
P_L(u_x \cdot u_y)= \omega_L Tr \bigl( {\sg W}_{x}^{(L)} \cdot {\sg
W}_{x}^{(L)} \bigr),
\la{elast03}
\ee
where ${\sg W}^{(L)}$ denote real multipole tensors of rank $L$
constructed in terms of components of $u_x$ and $u_y$,
respectively, and $\omega_L$ is an appropriate positive
normalization factor, so that
$Tr \bigl( {\sg W}_{x}^{(L)} \cdot {\sg W}_{x}^{(L)} \bigr)=
1/\omega_L$. Then, under the same hypotheses on $a(|x-y|)$ and
$g(|x-y|)$, one can prove absence or existence of an ordering
transition, respectively, for arbitrary $L$.
Even values of $L$ define lattice-gas models of nematic liquid
crystals,
and the result proven in Ref. \cite{r04} and for $d=3$ can be
similarly generalized.

%
%
In conclusion, we would like to point out that the present note
yields a partial answer to questions proposed in \cite{r01} as open
problems. We have shown that long-range interactions violating
conditions $(2.5)$ of Ref. \cite{r01} (cf. (\ref{e07}) ) can
produce long-range
{\it orientational order} in low-dimensional {\it lattice} models
of nematic liquid crystals. The problem of the possible existence
of long-range orientational order for {\it continuum} models
remains open.

Another open problem which can be formulated in the framework of
this {\it lattice} model (\ref{e03}) concerns existence of
long-range {\it positional order}, i.e.

$\lim_{|x-y|\rightarrow +\infty} [\langle n_x \cdot n_y \rangle -
\langle n_x \rangle \langle n_y \rangle ] > 0,$

when $g(|x-y|) \ge 0$, repulsion which crudely mimics
excluded-volume effects for nematic molecules.

\parindent 0cm
\section*{Acknowledgments}

The present text reached its final form during S.R.'s visit to
CPT-Luminy; the French CNRS is acknowledged for hospitality and
financial support.

\include{theosef2}
%
%
%
%
%
%
%
%
%
%

\end{document}